\documentclass[prd, twocolumn, superscriptaddress,floatfix, nofootinbib, preprintnumbers]{revtex4-2}
\pdfoutput=1
\usepackage[utf8]{inputenc}
\usepackage{hyperref}
\hypersetup{colorlinks=true,citecolor=blue}\usepackage{rotating}
\usepackage{array}
\usepackage{amsmath,amssymb,amstext}
\usepackage[normalem]{ulem}
\usepackage{slashed}
\usepackage{booktabs}
\usepackage[pdftex,table,dvipsnames]{xcolor}
\usepackage{multirow}
\usepackage{url}
\usepackage{color}
\usepackage{xspace}
\usepackage{comment}
\usepackage{enumitem}
\usepackage[caption=false]{subfig}
\usepackage{siunitx}
\usepackage{graphicx}
\usepackage{placeins}
\usepackage{layouts}
\usepackage{tikz}
\usetikzlibrary{shadings}           % GW: color gradients
\usetikzlibrary{arrows,calc,positioning,fit,matrix,shadows,chains,arrows,shapes,spy,fadings}
\usepackage{multirow}
\renewcommand{\vec}[1]{\mathbf{#1}}

\xdefinecolor{rwthblue}{rgb}{0,0.3294,0.6235}
\xdefinecolor{rwthlightblue}{rgb}{0.5569,0.7294,0.8980}

\xdefinecolor{rwthmagenta}{RGB}{227,0,102}
\xdefinecolor{rwthlightmagenta}{RGB}{241,158,177}

\xdefinecolor{rwthgrey}{RGB}{100,101,103}
\xdefinecolor{rwthlightgrey}{RGB}{156,158,159}

\xdefinecolor{rwthyellow}{RGB}{255,237,0}
\xdefinecolor{rwthlightyellow}{RGB}{255,245,155}

\xdefinecolor{rwthpetrol}{RGB}{0,97,101}
\xdefinecolor{rwthlightpetrol}{RGB}{125,164,167}

\xdefinecolor{rwthturquoise}{RGB}{0,152,161}
\xdefinecolor{rwthtlighturquoise}{RGB}{137,204,207}

\xdefinecolor{rwthgreen}{RGB}{87,171,39}
\xdefinecolor{rwthlightgreen}{RGB}{184,214,152}
\xdefinecolor{rwthlightlightgreen}{RGB}{221,235,206}

\xdefinecolor{rwthmaygreen}{RGB}{189,205,0}
\xdefinecolor{rwthlightmaygreen}{RGB}{189,205,0}

\xdefinecolor{rwthorange}{RGB}{246,168,0}
\xdefinecolor{rwthlightorange}{RGB}{253,212,143}

\xdefinecolor{rwthred}{RGB}{204,7,30}
\xdefinecolor{rwthlightred}{RGB}{230,150,121}
\xdefinecolor{rwthlightlightred}{RGB}{243,205,187}

\xdefinecolor{rwthbordeaux}{RGB}{161,16,53}
\xdefinecolor{rwthlightbordeaux}{RGB}{205,139,135}

\xdefinecolor{rwthviolet}{RGB}{97,33,88}
\xdefinecolor{rwthlightviolet}{RGB}{168,133,158}

\xdefinecolor{rwthpurple}{RGB}{122,111,172}
\xdefinecolor{rwthlightpurple}{RGB}{188,181,215}

\newcommand{\chidof}{$\chi^2/\text{dof}$}

\begin{document}

%\preprint{}

\title{Domain Adaptation against Background Sculpting in Anomaly Detection at the LHC}

\author{Vincent Benne}
\email{vincent.benne@rwth-aachen.de}
\affiliation{III. Physikalisches Institut, RWTH Aachen University,
D-52056 Aachen, Germany}

\author{Marie Hein}
\email{mahein@ethz.ch}
\affiliation{Institute for Theoretical Particle Physics and Cosmology, RWTH Aachen University, D-52056 Aachen, Germany}
\affiliation{Institute for Particle Physics and Astrophysics, ETH Z\"{u}rich, CH-8093 Z\"{u}rich, Switzerland}

\author{Michael Kr\"amer}
\email{mkraemer@physik.rwth-aachen.de}
\affiliation{Institute for Theoretical Particle Physics and Cosmology, RWTH Aachen University, D-52056 Aachen, Germany}

\author{Humberto Reyes-Gonz\'alez}
\email{humberto.reyes@rwth-aachen.de}
\affiliation{Institute for Theoretical Particle Physics and Cosmology, RWTH Aachen University, D-52056 Aachen, Germany}

\author{Philipp Soldin}
\email{soldin@physik.rwth-aachen.de}
\affiliation{III. Physikalisches Institut, RWTH Aachen University,
D-52056 Aachen, Germany}

\author{Christopher Wiebusch}
\email{wiebusch@physik.rwth-aachen.de}
\affiliation{III. Physikalisches Institut, RWTH Aachen University,
D-52056 Aachen, Germany}

\begin{abstract}
Weakly supervised anomaly detection has been shown to be an effective tool for model-agnostic searches for new physics, especially in the context of resonance searches. However, correlations between the anomaly score and the resonant mass can distort the background distribution after selecting on the anomaly score, complicating background estimation from the sidebands. To mitigate this background sculpting, we propose a domain-adaptation-based decorrelation of the anomaly score from the resonant mass. We study this approach using the LHC Olympics R\&D data set and several weakly supervised anomaly detection methods. We find that domain adaptation can substantially reduce background sculpting while largely preserving the anomaly detection performance. When correlations between the input features and the resonant mass degrade the original method's performance, domain adaptation can also recover sensitivity.
\end{abstract}

\maketitle

\section{Introduction}
\label{sec:Introduction}

Machine learning (ML)-based anomaly detection methods have gained significant traction in recent years. 
They can achieve high sensitivity to Beyond the Standard Model (BSM) signals with small cross sections, while remaining relatively agnostic to signal and background models
~\cite{Kasieczka:2021xcg, Aarrestad:2021oeb, Belis:2023mqs, hepmllivingreview}. 
These methods usually rely on the construction of an anomaly score
on which a selection can be based to enhance the signal significance. However, since these anomaly scores are derived from complex ML models, the potential for introducing biases or correlations into an analysis through this selection is much higher than for traditional selections based on single observables~\cite{Hallin:2022eoq}. 

Such correlations can be particularly problematic in resonance searches, where the standard procedure for determining the significance of a potential signal relies on the background prediction: For each signal mass hypothesis, a corresponding signal region (SR) is defined, and a fit is performed to the sidebands (SB), i.e., the largely signal-free regions on both sides of the SR. The fit is then used to predict the background in the SR, where the observed data are compared with this background prediction. However, a correlation between the anomaly score and the resonant mass can make the background estimate less reliable
because selecting on the anomaly score introduces a mass-dependent background efficiency. As a result, an initially smooth background distribution can acquire artificial structures after the selection, which may mimic or obscure a resonant signal. This problem is referred to as \emph{background sculpting}~\cite{Hallin:2022eoq}. In the past, anomaly detection analyses have addressed this problem in several ways, such as using quantile regression to retain similar statistics throughout the spectrum for autoencoders \cite{CMS:2024nsz}, or by defining the resonant mass so that its correlation with the features used to train the anomaly detector is limited \cite{ATLAS:2020iwa, ATLAS:2025obc}. Another solution was proposed in Ref.~\cite{Hallin:2022eoq} with the introduction of LaCATHODE, which is designed to mitigate background sculpting in weakly supervised searches. Alternative approaches to extract significances that do not rely on a fit have also been proposed \cite{DeSimone:2018efk,DAgnolo:2018cun,Letizia:2022xbe,Grosso:2024wjt, Das:2024eie}. However, these approaches present their own challenges, such as estimating systematic effects or determining the distribution of the test statistic.

In this work, we propose a novel solution to the background sculpting problem based on domain adaptation by backpropagation (DAB) \cite{ganin2015unsuperviseddomainadaptationbackpropagation}. Domain adaptation addresses situations where training and test data are drawn from different underlying distributions, referred to as domains. For example, one may wish to train a network on labeled simulation but apply it to measurement data. To do so successfully, the network should not be sensitive to any differences between the domains. Several domain-adaptation methods have been developed for this purpose, including DAB. DAB introduces a secondary task, called the domain task, which could, for example, distinguish simulated from measurement data. The model is trained to perform its main task while simultaneously preventing the learned representation from retaining information that distinguishes the two domains. At the LHC, DAB has been used in this way to reduce the effects of differences between data and Monte Carlo simulation~\cite{Baalouch:2019fhm, CMS:2025kgf, CMS:2022emx, CMS:2019dqq, CMS:2023jqi, Stein:2022nvf}. Related adversarial approaches have also been proposed to decorrelate anomaly detection from the jet mass \cite{Golling:2023juz} and studied to reduce the dependence of classifiers on systematic uncertainties \cite{Ghosh:2021roe}.

We study DAB against background sculpting in the context of weakly supervised anomaly detection \cite{Metodiev:2017vrx}.
Here, the anomaly detection task is formulated as a classification between the observed data in the signal region and a background template (BT) that is designed to approximate the background distribution in the signal region. In our specific case, the domain task is designed to reduce the correlation between the anomaly score and the resonant mass. We consider two domain tasks: a classification between the left and right sidebands and a regression of the resonant mass using sideband data. For our experimental setup and performance evaluation, we largely follow Ref.~\cite{Hallin:2022eoq}, allowing for a direct comparison with the background sculpting studies presented therein.

The remainder of this paper is organized as follows. In Sec.~\ref{sec:weaksupervision and domainadaptation}, we introduce the weakly supervised anomaly detection methods considered in this work, as well as the domain adaptation by backpropagation approach. In Sec.~\ref{sec:setup}, we describe the experimental setup and the metrics used to evaluate background sculpting and anomaly detection performance. We present our results in Sec.~\ref{sec:results} and conclude with a summary and outlook in Sec.~\ref{sec:conclusion}. The definition of the metric used to quantitatively evaluate sculpting severity is defined severity can be found in App.~\ref{app:chi2}. Details of the choice of the relative weight of the two loss terms used in DAB are provided in App.~\ref{app:tuning of lambda}. Additional visualizations of the results are provided in App.~\ref{app:additional spectra}.

\begin{figure*}
    \centering
    \includegraphics[width=.8\linewidth]{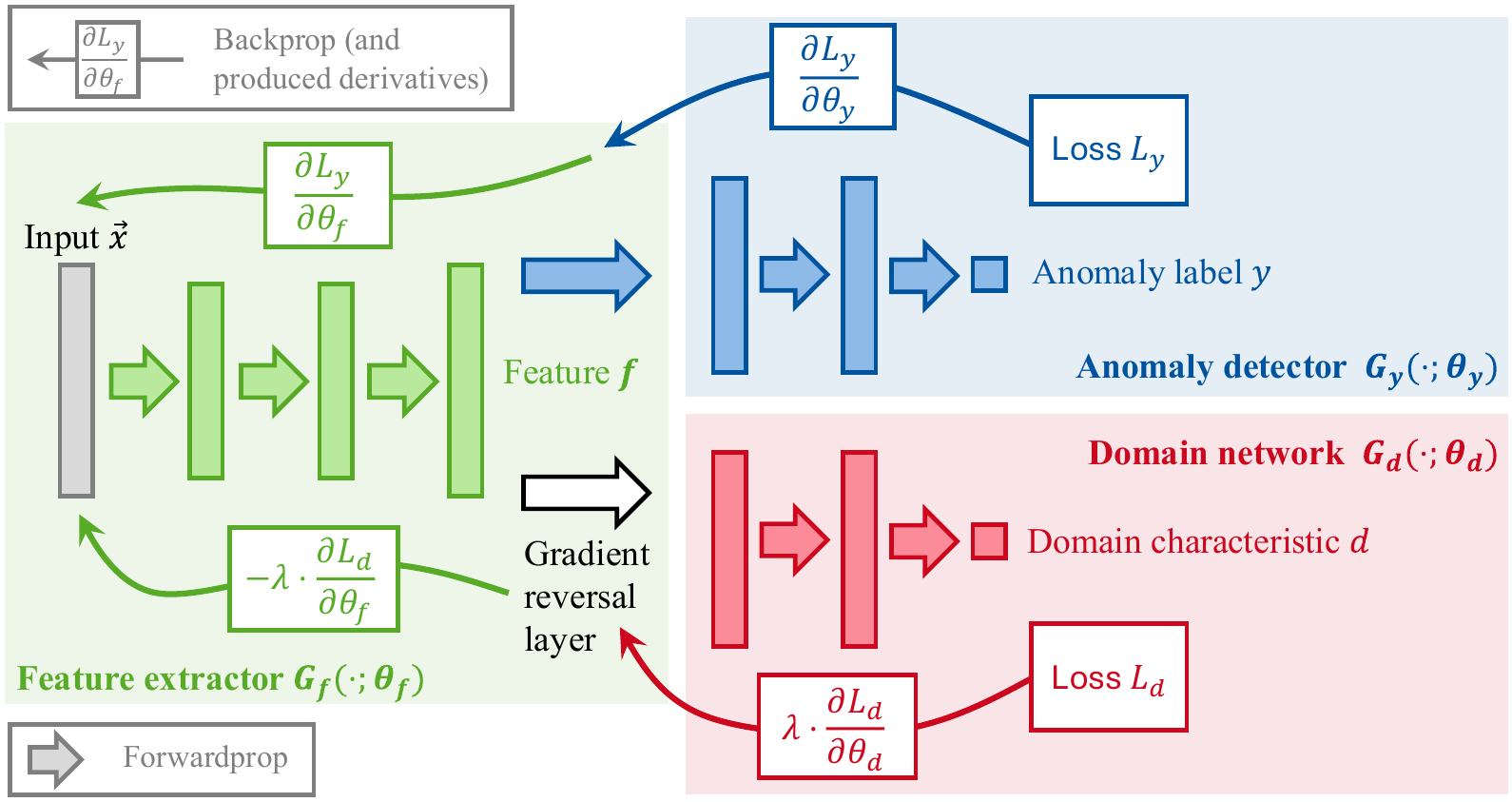}
    \caption{Illustration of domain adaptation by backpropagation (DAB), based on Ref.~\cite{ganin2015unsuperviseddomainadaptationbackpropagation}. The feature extractor is shown in green, the anomaly detector in blue, and the domain network in red.}
    \label{fig:domain adaptation}
\end{figure*}

\section{Method}
\label{sec:weaksupervision and domainadaptation}

\subsection{Weakly supervised anomaly detection}
In this work, we study weakly supervised anomaly detection in the context of a dijet resonance search using the LHC Olympics (LHCO) data set \cite{Kasieczka:2021xcg}. We use the dijet invariant mass $m_{JJ}$ as the resonant variable, while the anomaly detection task is based on additional jet features. Weakly supervised anomaly detection is framed as a classification task that distinguishes between events from a background template, representing a pure sample of background events in the SR, and events from the observed data in the SR. For a perfectly trained classifier and an ideal background template, the classifier score is equivalent to the optimal anomaly score according to the Neyman--Pearson lemma \cite{Neyman:1933wgr, Metodiev:2017vrx}. In practice, the main challenge lies in constructing an appropriate background template, for which a number of methods have been introduced \cite{Metodiev:2017vrx,Collins:2018epr,Collins:2019jip,Nachman:2020lpy,Andreassen:2020nkr,1815227,Hallin:2021wme,Raine:2022hht,Hallin:2022eoq,Golling:2022nkl,Golling:2023yjq,Das:2024fwo,Leigh:2024chm,Oleksiyuk:2025pmu}. Here, we consider four such methods to evaluate our approach to mitigating background sculpting.

First, we test our approach using the Idealized Anomaly Detector (IAD) \cite{Hallin:2021wme}, a benchmark method that uses an idealized background template made up entirely of SR background events. Second, we consider Classification Without Labels (CWoLa) Hunting \cite{Collins:2018epr, Collins:2019jip}, which uses background events from narrow sidebands around the signal region as the background template. Notably, CWoLa Hunting is particularly sensitive to correlations between the input features and the resonant mass, both in terms of anomaly detection performance and the degree of background sculpting. Third, we consider Classifying Anomalies THrough Outer Density Estimation (CATHODE) \cite{Hallin:2021wme}, which trains a conditional density estimator on the sidebands to learn $p(x|m_{JJ})$. The learned distribution is then interpolated into the SR to generate the background template. Finally, we compare our approach with LaCATHODE \cite{Hallin:2022eoq}, a previously introduced method for mitigating background sculpting. LaCATHODE uses a CATHODE-like density estimator but performs the classification in the density estimator's latent space. Events mapped to this latent space are expected to be decorrelated from $m_{JJ}$ and therefore less susceptible to background sculpting, as demonstrated in Ref.~\cite{Hallin:2022eoq}.

\subsection{Domain adaptation by backpropagation}

Domain adaptation by backpropagation was introduced in Ref.~\cite{ganin2015unsuperviseddomainadaptationbackpropagation} to learn representations that are simultaneously discriminative for the main learning task and invariant under changes between domains. DAB works by introducing an additional domain task alongside the main ML task. An illustration of this method is shown in Fig.~\ref{fig:domain adaptation}. In our application, the main task is anomaly detection and attempts to learn the weakly supervised anomaly label $y$, while the domain task attempts to predict the domain characteristic $d$. To obtain these two separate outputs, we split the standard MLP classifier used without domain adaptation into two parts: a ``feature extractor" $G_f$, shown in green, made up of the input layer and the first hidden layers, and the ``anomaly detector" $G_y$, shown in blue, made up of the remaining hidden layers and the output layer. A ``domain network" $G_d$, shown in red, is then attached to the feature extractor through a ``gradient reversal layer" and provides the domain output. This layer passes information without modification during the forward pass but reverses the gradient, i.e., multiplies it by $-1$, during the backward pass. The two losses are combined as
\begin{equation}
    L = L_y + \lambda \cdot L_d,
\end{equation}
where $L_y$ is the anomaly detection loss, $L_d$ is the domain loss, and the loss weight $\lambda$, which is an additional hyperparameter of the network that controls the influence of the domain task on the training. 
The anomaly-detector parameters $\theta_y$ are updated to minimize the anomaly detection loss $L_y$, while the domain-network parameters $\theta_d$ are updated to minimize the scaled domain loss $\lambda L_d$. Because the feature representation $\vec{f}$ is passed to both the anomaly detector and the domain network, the feature-extractor parameters $\theta_f$ are updated to 
minimize the anomaly detection loss while maximizing the domain loss. The maximization follows from the gradient reversal.

In our case, we consider two different domain tasks, both defined using sideband data. The first is a classification of background events from the left and right sidebands. Since both sidebands should consist almost exclusively of background events, the main difference between them is their invariant mass. The domain classifier can therefore distinguish between the two sidebands only to the extent that the learned representation retains information correlated with the invariant mass. Thus, preventing the domain classifier from distinguishing the two sidebands reduces the mass information retained in the learned representation and thereby promotes decorrelation from the invariant mass. Second, we employ a regression task for the resonant mass using the sideband data. Training the feature extractor to degrade this prediction reduces the information about $m_{JJ}$ retained in $\vec{f}$. The two domain tasks offer different advantages. When both the anomaly detection and domain tasks are classification tasks, their losses have a similar scale, which may facilitate the choice of the loss weight $\lambda$. On the other hand, the regression task is more directly aligned with our decorrelation goal. In practice, we find that both tasks perform similarly overall, with some differences that we discuss below.

\section{Experiment setup}
\label{sec:setup}

To evaluate the performance of the mitigation of background sculpting, we need to assess not only the degree of background sculpting but also its impact on signal sensitivity. For this, we train all anomaly detection methods in both background-only scenarios and with the inclusion of signal events. 

\subsection{Data set}
\label{sec: data set}
We perform our experiments using the R\&D data set~\cite{LHCOdataset} from the LHC Olympics 2020~\cite{Kasieczka:2021xcg} challenge, which has since become a common benchmark data set for anomaly detection. This data set contains 1M QCD dijet background events and 100k resonant $W'$ signal events decaying into an $X$ and $Y$ particle with masses $m_{W'}=3.5\,\text{TeV}$, $m_{X}=100\,\text{GeV}$ and $m_{Y}=500\,\text{GeV}$, respectively. $X$ and $Y$ each decay into two quarks, leading to a two-prong substructure. In addition to these events, we also make use of 612\,858 additional SR background events for a SR defined by $3.3\,\text{TeV} <m_{JJ}<3.7\,\text{TeV}$. These events were generated for Ref.~\cite{Hallin:2021wme} and published in Ref.~\cite{extraLHCOdataset}. Additionally, we use a set of 1M background events across the full invariant mass range published in Ref.~\cite{fanselow_2026_22685072}. All events were generated using \texttt{Pythia 8}~\cite{Sjostrand:2006za,Sjostrand:2007gs} and \texttt{Delphes 3.4.1}~\cite{deFavereau:2013fsa} with a leading-jet trigger of $p_{\mathrm{T}}>\SI{1.2}{\tera\electronvolt}$. We cluster jets using the anti-$k_{T}$ algorithm~\cite{Cacciari:2005hq} with a distance parameter of $R=1$ in \texttt{Fastjet}~\cite{Cacciari:2011ma} and use the two highest $p_T$ jets. 

As the resonant feature, we use the dijet invariant mass $m_{JJ}$. For the classification, we define two feature sets, one with only weak correlations with $m_{JJ}$ and one with stronger correlations, in order to study the effect of these correlations on background sculpting.
Both feature sets use the mass of the lighter jet $m_{J1}$, the mass difference of the two jets $\Delta m_J=m_{J_2}-m_{J_1}$, as well as both jets' 21-subjettiness ratios \cite{Thaler:2010tr, Thaler:2011gf}. These features are only very weakly correlated with $m_{JJ}$, and by using only these features, we obtain our ``Baseline'' feature set. If we include the angular distance between the jets $\Delta R$ in addition, we obtain our strongly correlated ``Baseline+$\Delta R$'' set.

We construct the data sets for our experiments as follows: As ``data'' we use all 1M background events from Ref.~\cite{LHCOdataset} and inject $N_\text{sig}$ signal events into the spectrum. The signal events are randomly selected for each run. The events from this set that fall into the SR, defined as described above, make up the signal-enriched sample for the weakly supervised classification in each of our methods. The background template varies by method and is described below. 
For the domain task, we use all events in the data sample outside the SR, i.e.\ the full sideband sample. To evaluate the anomaly detection performance, we construct a labeled test set from 340\,000 SR background events from Ref.~\cite{extraLHCOdataset} and 20\,000 SR signal events from Ref.~\cite{LHCOdataset}. To test the degree of sculpting, we use the 1M background events from Ref.~\cite{fanselow_2026_22685072}.

\subsection{Background template construction}

\subsubsection{Idealized Anomaly Detector (IAD)}
For the IAD, the background template consists of SR background events. For this, we use all but the 340\,000 test set events from Ref.~\cite{extraLHCOdataset}, resulting in a background template of approximately 272\,000 events. 

\subsubsection{CWoLa Hunting}
For CWoLa Hunting, our background template consists of events from narrow sidebands of $0.2\,\text{TeV}$ on each side of the SR.

\subsubsection{CATHODE}
For CATHODE, we train a conditional density estimator on the full sideband data, pre-processed using a logit transform to smooth sharp drop-offs. Our setup uses a conditional flow matching (CFM) density estimator based on Ref.~\cite{Das:2024fwo} using the code from Ref.~\cite{density_estimation}. We also use the hyperparameters from Ref.~\cite{Das:2024fwo}. For each signal injection, we train this setup ten times with random initializations and ensemble the resulting density estimators by combining their samples. For the background template, we use $4\cdot N_\text{data,SR}$ of these samples, where $N_\text{data,SR}$ is the number of SR data events, i.e., we oversample by a factor of four. 

\subsubsection{LaCATHODE}
For LaCATHODE, we use the same density estimators as trained for CATHODE. All data events are transported into the density estimator's latent space. We sample from the Gaussian in latent space to obtain a background template, again oversampling by a factor of four. This is the standard LaCATHODE construction: if the density estimator accurately models the background, background events are mapped to this common latent distribution. Transporting a pure SR background sample instead would require the idealized background information that LaCATHODE is intended to avoid. As different trainings may learn different orientations of the latent space, we cannot ensemble the density estimators for LaCATHODE. Therefore, we use a different density estimator for each anomaly detector training.

\subsection{Classifier implementations}
\label{sec: network implementations}

\subsubsection{Without domain adaptation}
Our classifier is a standard MLP implemented in \texttt{PyTorch}~\cite{Ansel_PyTorch_2_Faster_2024} based on Ref.~\cite{sk-cathode}. 
We use the hyperparameters from Ref.~\cite{Hallin:2021wme}. As such, the network consists of three hidden layers with 64 nodes each and ReLU activation. 
We train using the binary cross-entropy loss with the Adam optimizer \cite{Kingma:2014ad} with a learning rate of $10^{-3}$ for 100 epochs using a 50-50 training and validation split with a batch size of \num{128}. 
The final predictions are obtained by constructing an ensemble classifier from ten epochs selected according to the maximum ARGOS calculated on the validation set, a model-agnostic criterion for model selection defined in Ref.~\cite{Hein:2025uhj}. This differs from Ref.~\cite{Hallin:2021wme}, where epochs are selected based on the validation loss; Ref.~\cite{Hein:2025uhj} found the ARGOS-based selection to provide a slight performance gain. We restrict the selection to the second half of the training for consistency with the domain-adaptation implementation, which requires an initial warm-up period, as described in the next section. We further ensemble a total of five such classifiers. For all ensembles, we use the mean prediction for each event.

\subsubsection{With domain adaptation}

\begin{table}[t]
    \centering
    \begin{tabular}{l|c|c|c|c}
         Domain Task&  \multicolumn{2}{c|}{Classification} & \multicolumn{2}{c}{Regression}\\[1mm]
         Input Set & Baseline & $+\Delta R$ & Baseline & $+\Delta R$ \\ \hline
         IAD & 0.1 & 0.1 & 0.2 & 0.2 \\
         CATHODE & 0.1 & 0.1 & 0.2 & 0.2 \\
         CWoLa Hunting & 0.4 & 1.5 & 0.2 & 2 \\
    \end{tabular}
    \caption{Loss weight $\lambda$ as defined in Fig.~\ref{fig:domain adaptation} used to obtain the results in this paper.}
    \label{tab:selected weights}
\end{table}

For the implementation using domain adaptation, we largely retain the training setup described above and modify the network structure. Our feature extractor consists of two hidden layers of 64 nodes each, with ReLU activation, where the output of the second layer defines the feature vector $\vec f$. The anomaly detector $G_y$ then uses another hidden layer of 64 nodes plus the output layer, such that the total anomaly detection network is of the same size as in the case without domain adaptation; see Fig.~\ref{fig:domain adaptation} for reference. The domain network $G_d$ uses a single hidden layer of 64 nodes with ReLU activation, as well as a single-node output layer. The anomaly and domain losses are evaluated on separate batches from their respective data sets. For both data sets, we use a 50–50 training-validation split and a batch size of 128. We use the same number of batches for the two tasks. Because the domain-task data set is larger, we subsample it in each epoch.

The full network architecture introduces an additional hyperparameter $\lambda$ that scales the domain adaptation loss term. For the results presented here, we performed a limited tuning of this parameter, described in App.~\ref{app:tuning of lambda}. The resulting values of $\lambda$ used to obtain the results below can be found in Tab.~\ref{tab:selected weights}. 

\subsection{Evaluating the performance}
\label{sec: performance evaluation}

To study background sculpting, we perform background-only runs and apply selections on the anomaly score. For each setup, we choose the anomaly-score threshold such that it retains $1\%$ of the events in the SR and show the spectrum obtained by applying this threshold over the full mass range. This provides an intuitive illustration of background sculpting. 
We additionally show the spectra from all ten runs in App.~\ref{app:additional spectra} to illustrate the variation between runs and ensure that our conclusions are not based on statistical fluctuations of individual example spectra. 
To compare the setups quantitatively, we calculate the binned \chidof\ metric introduced in Ref.~\cite{Hallin:2022eoq}.
This metric compares the normalized $m_{JJ}$ distributions before and after the anomaly-score selection, using the statistical uncertainty expected for a random selection at the same efficiency. 
Values of \chidof\ close to unity indicate that changes in the shape of the background distribution are compatible with statistical fluctuations, while larger values indicate background sculpting. 
A detailed definition of the metric is provided in App.~\ref{app:chi2}.

To evaluate the anomaly detection performance, we perform runs with varying signal injections $N_{sig}$ and evaluate the significance improvement characteristic (SIC) defined as 
\begin{equation}
    \text{SIC}=\frac{\epsilon_S}{\sqrt{\epsilon_B}},
\end{equation}
where $\epsilon_{S/B}$ is the signal/background efficiency of the respective selection cut. 
The SIC approximates the improvement in the Poisson significance in the Gaussian limit. We apply a statistical cutoff to the SIC corresponding to a maximum 20\% statistical error on the background estimate. We generally consider the maximum SIC at different signal injections to evaluate the sensitivity of the anomaly detector.

\begin{figure*}[p]
    \centering
    \includegraphics{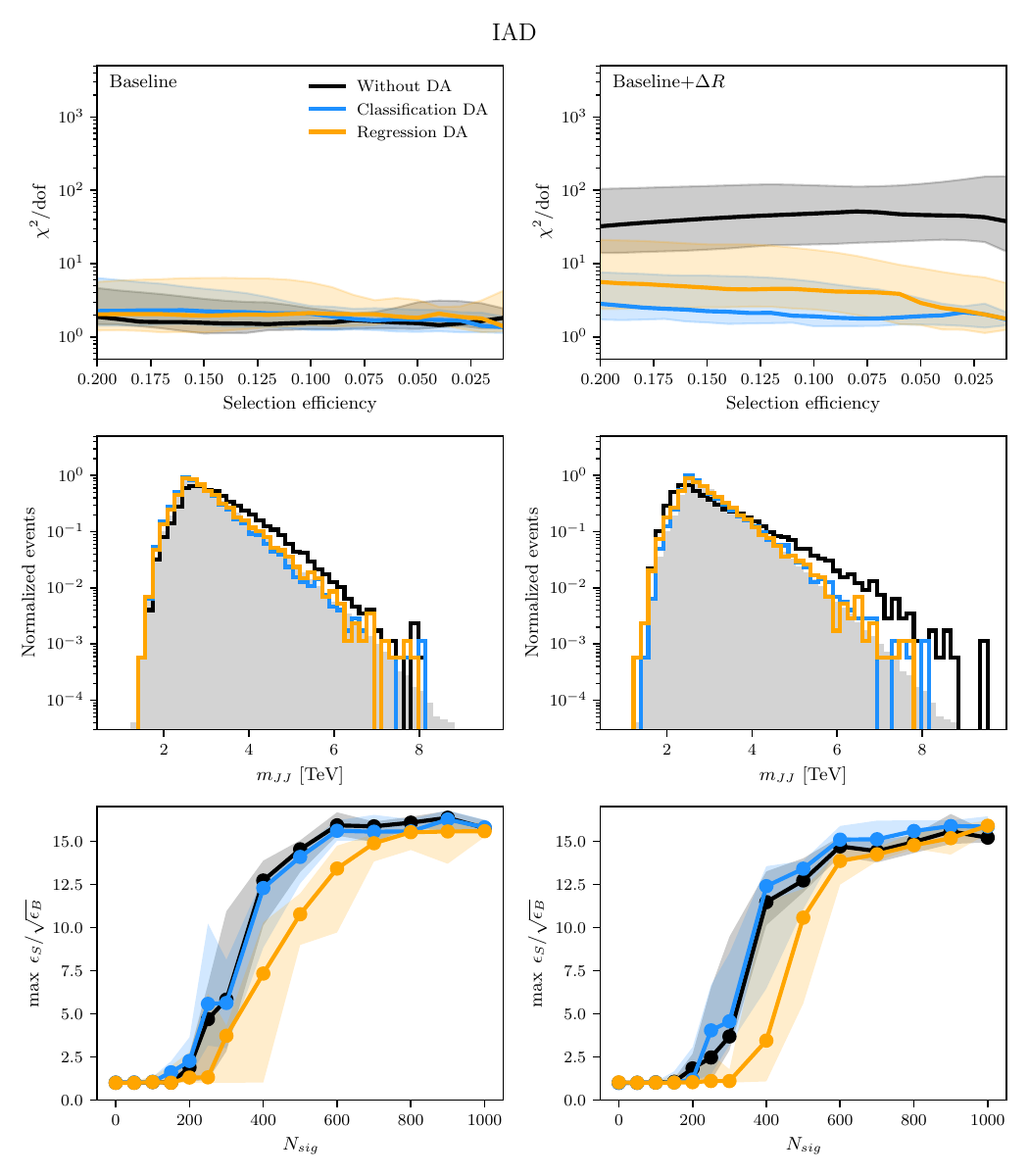}
    \caption{Sculpting severity and anomaly detection performance for the IAD using the ``Baseline" (left) and ``Baseline+$\Delta R$" (right) feature sets. Results are shown without domain adaptation (black) and with domain adaptation using the classification (blue) or regression (orange) task. We show the sculpting severity both as the binned $\chi^2$ per degree of freedom metric described in Sec.~\ref{sec: performance evaluation} (top) and as an example spectrum (middle). We show the anomaly detection performance as the maximum SIC for different signal injections $N_\text{sig}$ (bottom). Both the $\chi^2$ metric and the maximum SIC are shown as the median and 68\% quantile interval of ten runs.}
    \label{fig:IAD results}
\end{figure*}

\section{Experiments}
\label{sec:results}

We now turn to our results, which we structure by the anomaly detection method, starting from the Idealized Anomaly Detector as a controlled benchmark, then considering CATHODE as a realistic weakly supervised setup where we also compare our approach to LaCATHODE, before finally discussing CWoLa Hunting, where correlations between the input features and the resonant mass pose a particularly severe challenge.

\subsection{Idealized Anomaly Detector (IAD)}

We start with the IAD results shown in Fig.~\ref{fig:IAD results}, which provide a controlled test of the impact of input feature correlations on background sculpting. Without domain adaptation, the Baseline feature set shows only a small degree of background sculpting, both in the \chidof\ and in the example spectrum. The example spectrum without DA here looks slightly more distorted than those with DA. This is a combination of a slightly unfavorable random draw (see the full set of spectra in Fig.~\ref{fig:IAD spectra} in App.~\ref{app:additional spectra}) and the fact that we show spectra at a 1\% selection efficiency, which is located at the rightmost edge of the \chidof plot.
Nevertheless, all spectra remain sufficiently smooth to allow for a background fit. 
The situation changes considerably when the additional feature $\Delta R$ is included. For the Baseline+$\Delta R$ feature set, the \chidof\ increases by more than an order of magnitude, indicating substantially stronger background sculpting. The example spectrum shown in Fig.~\ref{fig:IAD results} represents a relatively mild case; several of the other runs show pronounced dips or bumps in the SR as can be seen in Fig.~\ref{fig:IAD spectra} in App.~\ref{app:additional spectra}. In these cases, the spectrum is no longer monotonically falling, making a standard fit-based bump hunt difficult. For both feature sets, the IAD nevertheless shows strong anomaly-detection performance, reaching a maximum SIC of about 16 for large-signal injections. Towards smaller signal injections, the SIC drops sharply around $N_\text{sig}\sim 300$, as is characteristic of weakly supervised anomaly detection (see, e.g., Ref.~\cite{Hallin:2021wme}).

Applying domain adaptation substantially reduces the background sculpting. 
Both the classification and regression domain tasks reduce the correlation between the anomaly score and the dijet mass, with the improvement being particularly pronounced for the Baseline+$\Delta R$ feature set. 
The classification task performs particularly well: for both feature sets, the \chidof\ is reduced to values close to unity, while the anomaly detection performance is essentially preserved.
The regression task also strongly reduces the sculpting, but shows some loss in signal sensitivity, with the rise of the maximum SIC shifted towards larger signal injections for both feature sets. 
This loss may be reduced by a more extensive optimization of the domain-loss weight $\lambda $. 
For the Baseline+$\Delta R$ feature set, the regression task also leaves somewhat more residual sculpting than the classification task. Thus, for the IAD setup considered here, the classification task provides the best overall compromise between background decorrelation and signal sensitivity. The spectra obtained with domain adaptation show a similar behavior across all ten runs (see Fig.~\ref{fig:IAD spectra} in App.~\ref{app:additional spectra}), with the spectra shown in Fig.~\ref{fig:IAD results} providing representative examples. 

\subsection{CATHODE}

\begin{figure*}[p]
    \centering
    \includegraphics{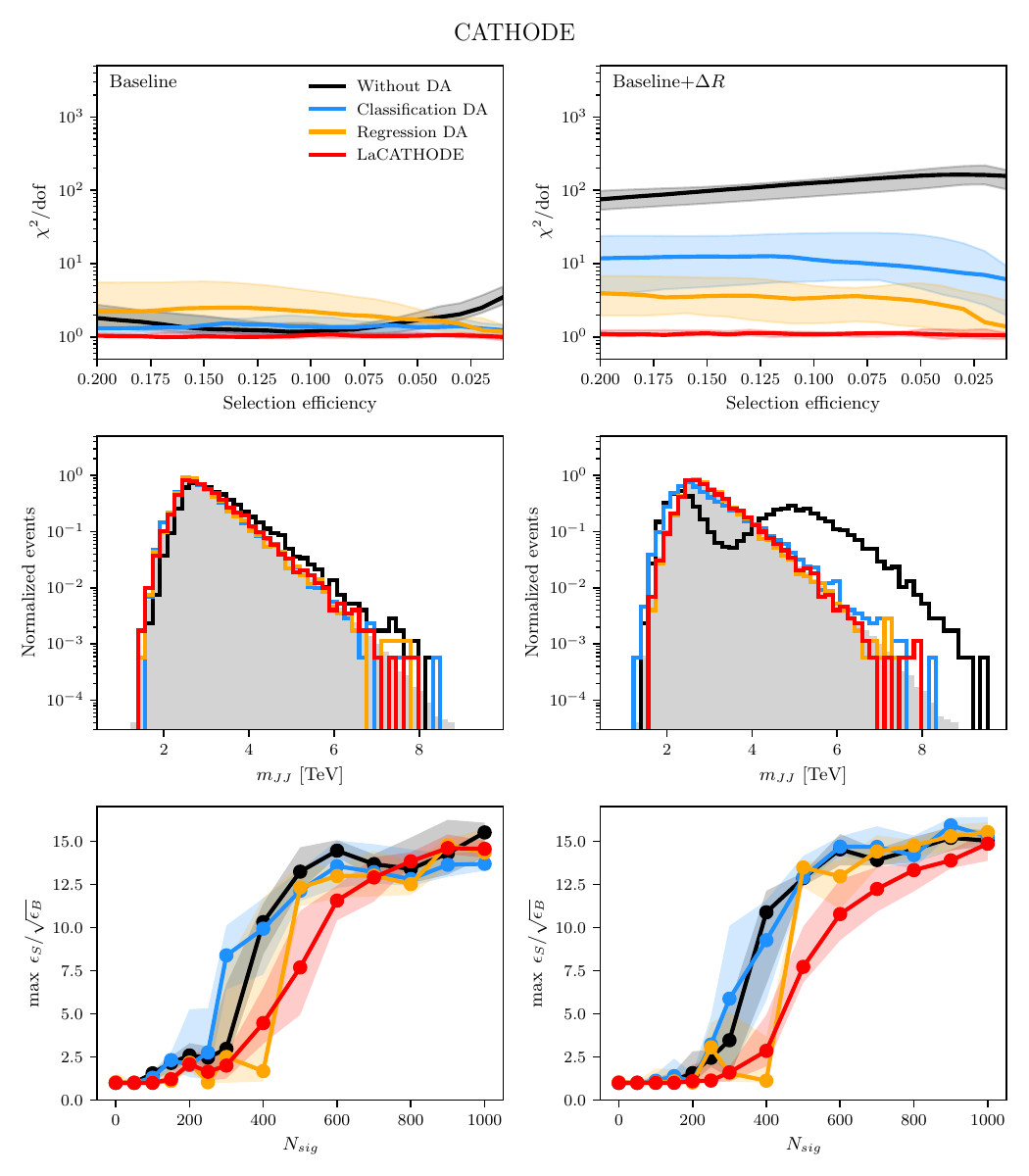}
    \caption{Sculpting severity and anomaly detection performance for CATHODE using the ``Baseline'' (left) and ``Baseline+$\Delta R$'' (right) features without (black) and with domain adaptation, both using the classification (blue) and regression (orange) task. Additionally, we show the performance of LaCATHODE (red). We show the sculpting severity both as the binned $\chi^2$ per degree of freedom metric described in Sec.~\ref{sec: performance evaluation} (top) and a single example spectrum (middle). We show the anomaly detection performance as the maximum SIC for different signal injections $N_\text{sig}$ (bottom). Both the $\chi^2$ metric and the maximum SIC are shown as the median and 68\% quantile interval of ten runs.}
    \label{fig:cathode results}
\end{figure*}

We next consider the CATHODE results shown in Fig.~\ref{fig:cathode results}. Compared with the IAD case, CATHODE includes the additional challenge of constructing the background template from sideband data. Nevertheless, the overall behavior is very similar to the IAD case, with the dominant effect again being the correlation introduced by the $\Delta R$ feature. In particular, the inclusion of $\Delta R$ introduces a strong correlation between the anomaly score and the resonant mass, leading to pronounced background sculpting in the absence of domain adaptation. For the Baseline+$\Delta R$ feature set, the selected background spectrum develops a clear non-monotonic structure, which would make a standard background fit difficult. This effect is observed consistently across the different runs, rather than being a feature of a single example spectrum (see Fig.~\ref{fig:cathode spectra} in App.~\ref{app:additional spectra}). Both domain adaptation tasks substantially reduce this sculpting. While the resulting \chidof\ values are not always close to unity, the corresponding spectra are smoothly falling and qualitatively very different from the strongly distorted spectra obtained without domain adaptation. In a realistic analysis, the relevant question is whether the remaining deviations can be described by a suitable background fit, for which the improvement compared with the setup without domain adaptation is substantial.

The anomaly detection performance is largely preserved when applying domain adaptation. As in the IAD case, the classification domain task provides the best overall compromise between sculpting reduction and signal sensitivity. The regression task also reduces the dependence on $m_{JJ}$, but shows a somewhat larger loss in sensitivity for some configurations. A more extensive optimization of the domain-loss weight may further improve this behavior.

We additionally compare the domain adaptation approach with LaCATHODE, which was specifically designed to mitigate background sculpting in CATHODE-based anomaly detection. As expected, LaCATHODE strongly reduces the sculpting for both feature sets and provides the smoothest background spectra. However, the anomaly detection performance is somewhat reduced compared with the original CATHODE setup and with our domain adaptation approach. This behavior is consistent with the observations in Ref.~\cite{Hallin:2022eoq}. The domain adaptation approach therefore provides a complementary strategy: while LaCATHODE aims for an explicitly decorrelated latent representation, domain adaptation preserves the original classification setup and enforces decorrelation with respect to the learned features relevant for the anomaly score. This allows the anomaly detection performance to be maintained more closely in the examples studied here.

Finally, we note that the comparison between CATHODE and LaCATHODE is affected by a technical difference in the implementation. For CATHODE, we use an ensemble of density estimators, while for LaCATHODE such an ensemble cannot be constructed in the same way because different density estimators correspond to different orientations of the latent space. Nevertheless, all domain adaptation spectra show a similar behavior across the ten runs, indicating that the observed behavior is not driven by a particularly favorable example.

\begin{figure*}[p]
    \centering
    \includegraphics{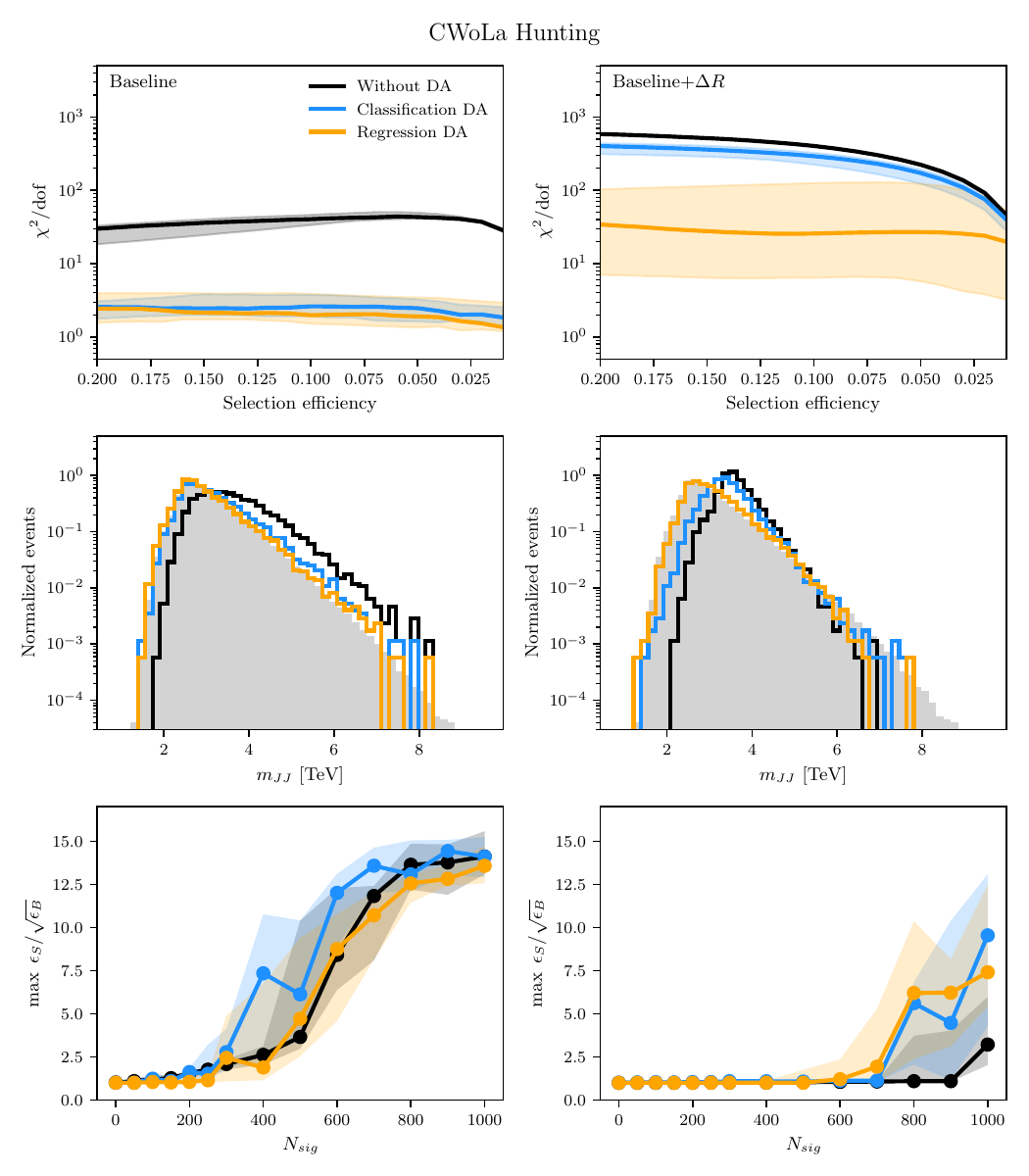}
    \caption{Sculpting severity and anomaly detection performance for CWoLa Hunting using the ``Baseline" (left) and ``Baseline+$\Delta R$" (right) feature sets. Results are shown without domain adaptation (black) and with domain adaptation using the classification (blue) or regression (orange) task. We show the sculpting severity both as the binned $\chi^2$ per degree of freedom metric described in Sec.~\ref{sec: performance evaluation} (top) and as an example spectrum (middle). We show the anomaly detection performance as the maximum SIC for different signal injections $N_\text{sig}$ (bottom). Both the $\chi^2$ metric and the maximum SIC are shown as the median and 68\% quantile interval of ten runs.}
    \label{fig:cwola results}
\end{figure*}

\subsection{CWoLa Hunting}
We finally consider CWoLa Hunting, which is particularly sensitive to correlations between the input features and the resonant mass. This is already visible for the Baseline feature set without domain adaptation in Fig.~\ref{fig:cwola results}. Since CWoLa Hunting classifies events in the SR against events from the sidebands, correlations with $m_{JJ}$ allow the classifier to distinguish the two samples even in the absence of a signal. As a result, the anomaly-score selection preferentially selects events in the SR and produces an artificial bump in the background spectrum. This effect is already present for the Baseline features and becomes considerably stronger when $\Delta R$ is included.

For the Baseline feature set, domain adaptation is very effective in reducing this sculpting. Both domain tasks lead to spectra that are approximately smooth. At the same time, the anomaly detection performance is preserved and even slightly improved for the classification task. Thus, in this case domain adaptation substantially reduces the unwanted dependence on the resonant mass without reducing the sensitivity to an anomalous signal.

The Baseline+$\Delta R$ feature set provides a considerably more challenging test. The strong correlation of $\Delta R$ with $m_{JJ}$ violates the basic assumption underlying CWoLa Hunting that, in the absence of a signal, the classifier should not be able to distinguish the SR from the sidebands. Accordingly, CWoLa Hunting without domain adaptation shows both severe background sculpting and essentially no anomaly detection sensitivity. This setup is therefore not one that would be used in a realistic CWoLa analysis, but provides a useful test of the ability of domain adaptation to recover sensitivity when the assumptions of CWoLa Hunting are violated. The classification domain task is not sufficient to remove the sculpting in this case; we find similar behavior for domain-loss weights between $\lambda=0.02$ and $5$. The regression task performs considerably better, removing the artificial SR peak in five out of ten runs (see Fig.~\ref{fig:cwola spectra} in App.~\ref{app:additional spectra}). Moreover, both domain adaptation tasks recover some anomaly detection sensitivity compared with the setup without domain adaptation. 
An interesting question for future investigations is to understand the better performance and whether it is related to its more direct constraint on the information about $m_{JJ}$ retained in the learned representation.

\section{Conclusion}
\label{sec:conclusion}
In resonance-based searches for new physics, correlations between the anomaly score and the resonant variable can distort the background distribution after a selection on the anomaly score. This complicates the background estimation from sidebands. In this paper, we have introduced a method for mitigating such background sculpting based on domain adaptation by backpropagation. We consider two domain tasks, both trained on sideband data: a classification between the left and right sidebands and a regression of the resonant variable. We study the approach using feature sets with weak and strong correlations to the dijet mass for the IAD, CATHODE, and CWoLa Hunting methods.

Overall, domain adaptation substantially reduces background sculpting while largely preserving the anomaly detection performance, although the results depend on the domain task and the anomaly detection method. For the IAD and CATHODE, the classification task provides the best compromise between sculpting reduction and signal sensitivity. The regression task also reduces sculpting, but leads to some loss in sensitivity. For CWoLa Hunting with the Baseline feature set, both domain tasks substantially reduce sculpting while preserving the anomaly detection performance. For the more challenging Baseline+$\Delta R$ feature set, where CWoLa Hunting without domain adaptation shows severe background sculpting and essentially no anomaly detection sensitivity, both domain tasks recover some sensitivity, while the regression task provides a stronger reduction in background sculpting.

The main advantage of our domain-adaptation-based approach is its flexibility. While LaCATHODE mitigates background sculpting by performing the CATHODE classification in an approximately decorrelated latent space~\cite{Hallin:2022eoq}, our approach acts directly on the learned feature representation and is not tied to a particular background template construction. The two approaches therefore provide complementary strategies for mitigating background sculpting. In particular, domain adaptation can in principle be combined with different background template generation methods, including recent proposals such as TRANSIT~\cite{Oleksiyuk:2025pmu} and RAD-OT~\cite{Leigh:2024chm}. Other neural-network-based anomaly detectors, such as autoencoders, could likewise be modified to include domain-adaptation-based sculpting mitigation.

The method itself presents two main practical challenges. 
First, it introduces the additional hyperparameter, the loss weight $\lambda$, which controls the trade-off between reducing background sculpting and preserving anomaly detection performance. 
In this study, we selected $\lambda$ through a limited optimization considering both sculpting mitigation and signal sensitivity (see App.~\ref{app:tuning of lambda}).
Such a procedure would not be possible in this form for a truly model-agnostic analysis, and developing a robust tuning strategy that does not rely on a specific signal model remains an important challenge. Second, domain adaptation by backpropagation requires a differentiable model trained through gradient-based optimization and therefore cannot be applied directly to boosted decision trees, which are common for weakly supervised anomaly detection using high-level features~\cite{Finke:2023ltw,Freytsis:2023cjr}.

The scope of the present study is restricted in two respects. First, we considered the sensitivity to a single signal benchmark only. Since the decorrelation is learned using sideband background events, we expect the method to reduce background sculpting for a broad range of resonant signals. However, the preservation of signal sensitivity was not studied for other signal models. Second, rather than performing complete sideband fits to the selected background spectra, we quantified background sculpting using a binned \chidof\ metric comparing the spectra before and after the anomaly-score selection. 
This metric provides a quantitative measure of background sculpting and facilitates comparison with previous studies~\cite{Hallin:2022eoq}, but does not determine the impact on the background prediction and the final statistical inference in a bump hunt.

Several directions for future work follow from these results. Automatic loss-weighting methods could reduce or eliminate the need for an application-specific choice of $\lambda$; our initial studies using homoscedastic loss weighting~\cite{1705.07115} showed promising results, although further optimization is required. The open questions identified above should be addressed through systematic evaluations using a broader range of signal models, more realistic data sets, and complete sideband fits. Further studies could explore extensions to neural-network architectures designed for tabular data, as well as alternative decorrelation strategies suitable for BDTs. More broadly, our results show that domain adaptation provides a flexible way of controlling background sculpting while retaining the advantages of weakly supervised anomaly detection, and may thereby broaden the range of features and background-template constructions that can be used in resonance searches.

\begin{figure*}[t]
    \centering
    \includegraphics{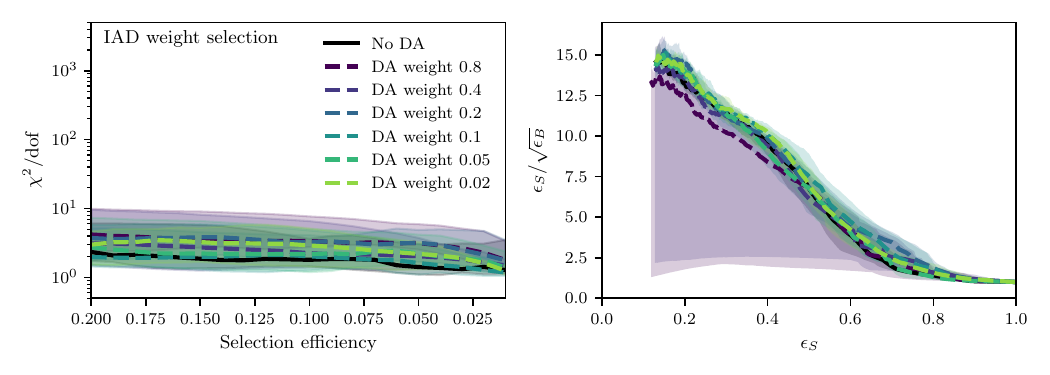}
    \caption{Sculpting severity and anomaly detection performance for IAD using the ``Baseline" feature set. Results are shown without domain adaptation (black) and with domain adaptation using the classification task with different loss weights $\lambda$ between 0.02 and 0.8 (colored). We show the sculpting severity as the binned $\chi^2$ per degree of freedom metric described in Sec.~\ref{sec: performance evaluation} (left). We show the anomaly detection performance as a SIC curve for $N_\text{sig}=1000$ (right). All results are shown as the median and 68\% quantile interval of ten runs.}
    \label{fig:weight scan}
\end{figure*}

\section*{Acknowledgements}
We thank Alexander Mück for helpful discussions. We also thank Alexander Mück and David Shih for valuable comments on the manuscript. 
MH is supported by the Deutsche Forschungsgemeinschaft (DFG, German Research Foundation) under grant 400140256 - GRK 2497: 
The physics of the heaviest particles at the Large Hadron Collider. 
The research of MK and HRG is supported by the SciFM consortium (05D25PA5) funded by the German Federal Ministry of Research, Technology, and Space (BMFTR) in the ErUM-Data action plan and the DFG under grant 396021762 - TRR 257: Particle Physics Phenomenology after the Higgs Discovery. 
PS is supported by the ERUM-ARIA project (05D25PA4) funded by the German Federal Ministry of Research, Technology, and Space (BMFTR) in the ErUM-Data action plan. This work has benefited from activities carried out within the framework of the COST Action MLQC4FC (CA24146), supported by COST (European Cooperation in Science and Technology). Computations were performed with computing resources granted by RWTH Aachen University under projects thes2064, rwth0934 and p0025242.

\section*{Code}
The code for this paper can be found in the GitHub repository~\cite{domainadaptationgithub}.

\appendix

\section{Sculpting metric}
\label{app:chi2}
For the $\chi^2$-metric as also used in Ref.~\cite{Hallin:2022eoq}, quantile-based binning is applied to the $m_{JJ}$ spectrum before the cut with $n_\text{bins}=300$. The number of events per bin normalized to unit area is then given by 
\begin{equation}
    n^\text{full} = a \frac{N_\text{tot}}{n_\text{bins}}
\end{equation}
with the normalization factor $a$. The same binning is then applied to the spectrum after the cut to obtain the number of selected events in bin $i$ normalized to unit area
\begin{equation}
    n_i^\text{sel} = \frac{a}{\varepsilon_\text{sel}}N_i^\text{sel},
\end{equation}
where $\varepsilon_\text{sel}$ is the selection efficiency and $N_i^\text{sel}$ is the number of selected events. Using these counts, the $\chi^2$ is then defined as
\begin{equation}
    \chi^2 = \sum_i \frac{(n^\text{full}-n_i^\text{sel})^2}{a n^\text{full}}\cdot \frac{\varepsilon_\text{sel}}{1-\varepsilon_\text{sel}}.
\end{equation}

\section{Weight tuning}
\label{app:tuning of lambda}

The correct weighting of the two loss terms in our domain-adaptation-based sculpting removal is crucial. If the domain loss is weighted too strongly, it may remove sculpting trivially by removing all information, including information useful for the anomaly detection task. If the anomaly detection loss is weighted too strongly, sculpting is not removed. To make an informed choice of the loss weight, we performed limited tuning by scanning over a range of six values for $\lambda$. For each value of $\lambda$, we used a single classifier training rather than the ensemble of five classifiers described in Sec.~\ref{sec: network implementations} and considered both the \chidof\ for the background-only case and the SIC curve at a nominal signal value of $N_\text{sig}=1000$. We show an example of these scans in Fig.~\ref{fig:weight scan} for the IAD using Baseline features in the classification domain adaptation task. We see that too large a weight leads to unstable anomaly detection performance (right panel), with the large error band indicating that some runs are completely unable to detect the signal. The sculpting mitigation, on the other hand, depends only very slightly on the chosen weight. Generally, these trends persisted throughout our tuning of $\lambda$, though CWoLa Hunting depended more strongly on this choice and required a larger value for $\lambda$, particularly when $\Delta R$ was included.

Based on our scans, we then selected the $\lambda$ values. Here, we chose to prioritize the anomaly detection performance over the sculpting performance and selected values significantly below those for which instability was observed. For the case shown in Fig.~\ref{fig:weight scan}, a value of 0.1 was chosen.

The tuning performed here considered both signal sensitivity and sculpting mitigation, which would not be possible in this form in a truly model-agnostic analysis. Simulation-based studies using example signals could be used to tune $\lambda$ for an analysis, though the use of specific signal models for tuning limits the degree of model agnosticism. As we generally observed limited dependence on $\lambda$, we believe that tuning on background-only simulations may be sufficient. Here, the smallest value of $\lambda$ that sufficiently suppresses sculpting should be chosen to limit the impact on signal sensitivity. Additionally, data-driven optimization methods, such as those introduced in Ref.~\cite{Hein:2025uhj}, could be used to optimize signal sensitivity directly on data.

An alternative is to optimize the loss weight automatically using the method of Ref.~\cite{1705.07115}. The homoscedastic loss weighting introduced there aims to weight multiple tasks by introducing a learned uncertainty parameter. Initial studies using this methodology for our domain-adaptation-based sculpting mitigation showed promising, although slightly unstable, results. A particular difficulty with this method is that we generally require the sensitivity to remain high, while we may be able to tolerate a small residual correlation with the resonant mass. Such a specific prioritization of tasks is difficult to build into the learned weighting. Nevertheless, we believe homoscedastic loss weighting to be a promising approach, as it could eliminate the loss-weight parameter. However, further optimization would be required to make it work reliably.

\section{Additional Mass Spectra}
\label{app:additional spectra}

\begin{figure*}[p]
    \centering
    \includegraphics{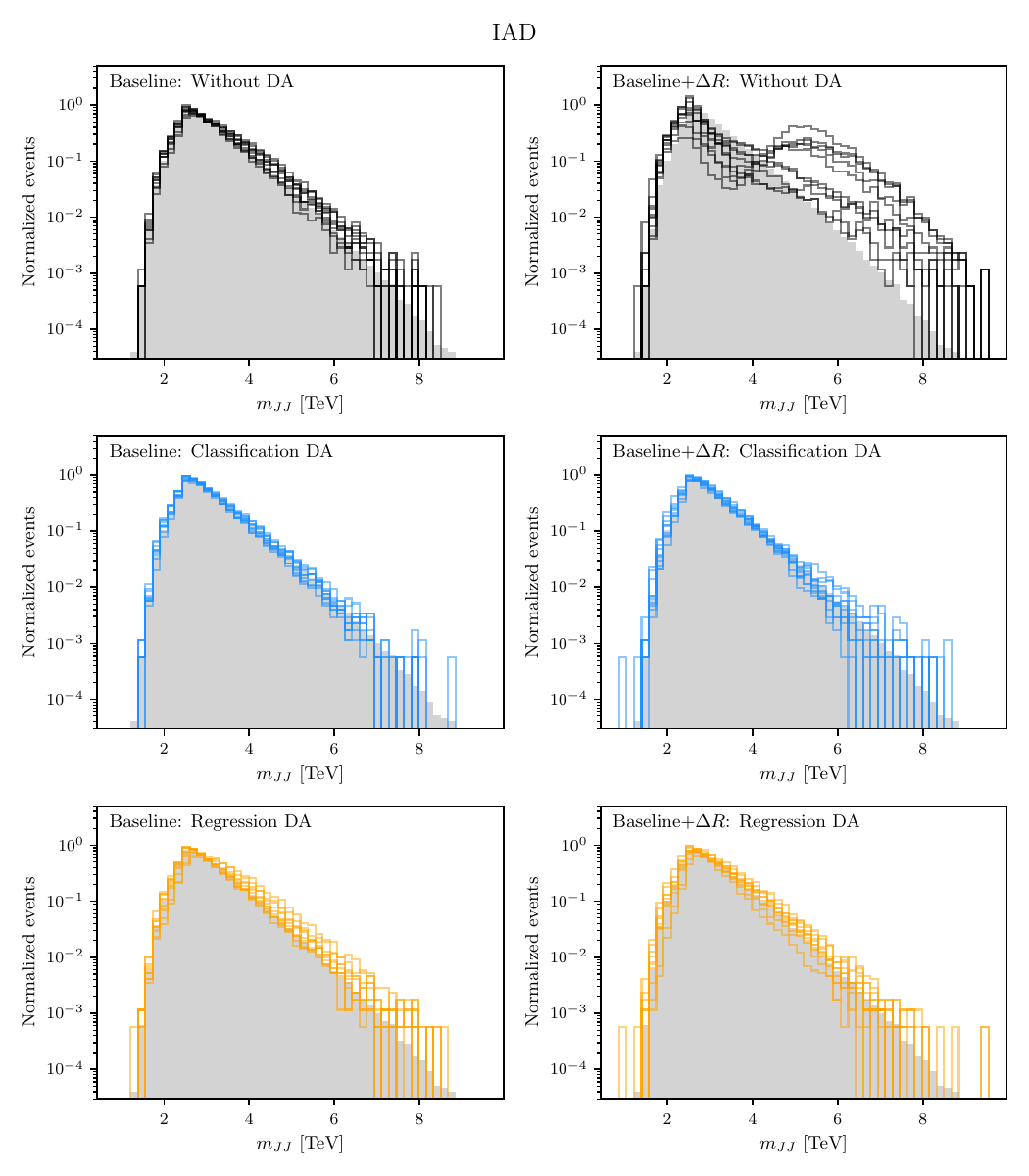}
    \caption{Full set of spectra for ten runs for the IAD using the ``Baseline" (left) and ``Baseline+$\Delta R$" (right) feature sets. Results are shown without domain adaptation (top) and with domain adaptation using the classification (middle) or regression (bottom) task.}
    \label{fig:IAD spectra}
\end{figure*}

\begin{figure*}[p]
    \centering
    \includegraphics{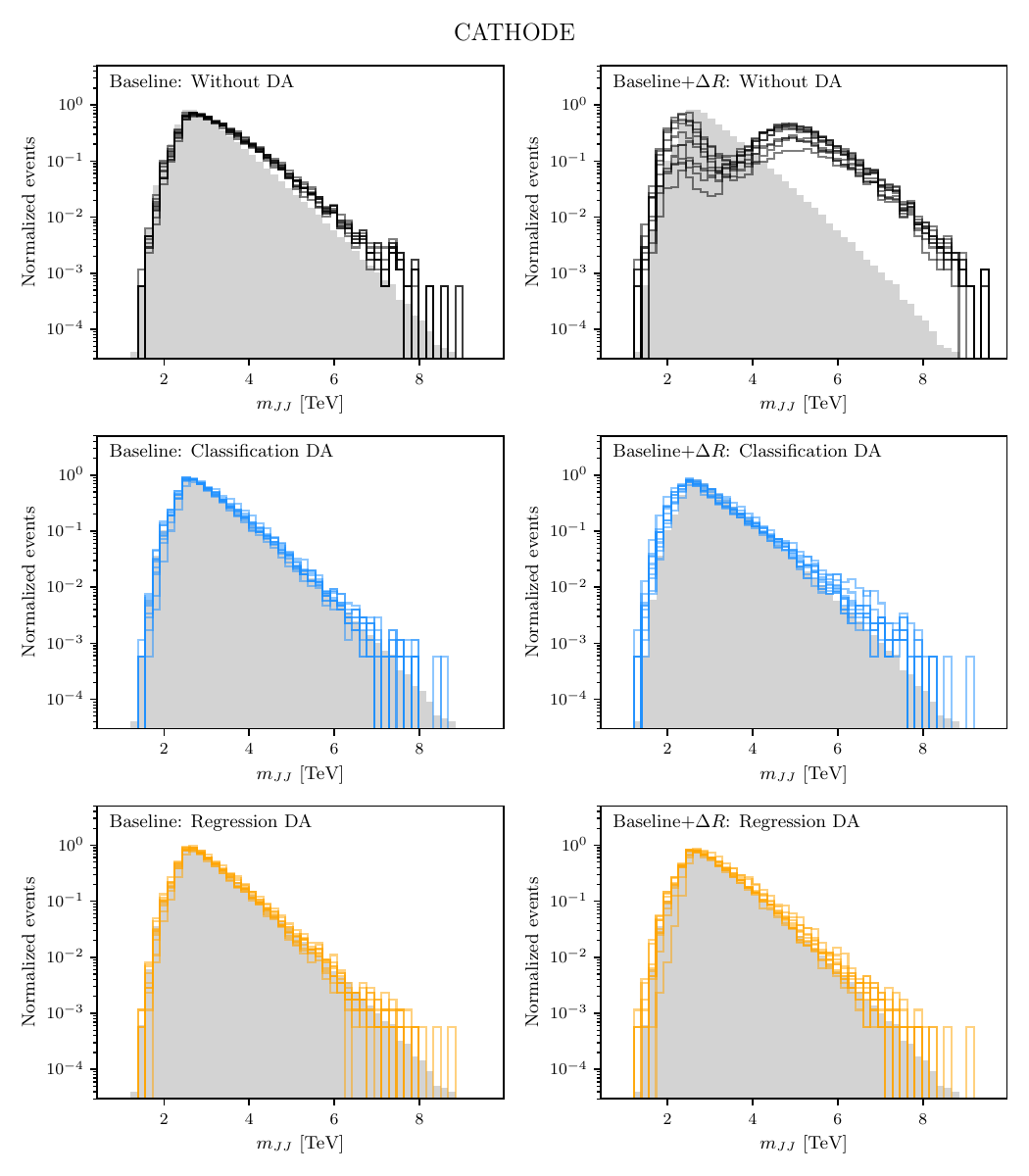}
    \caption{Full set of spectra for ten runs for CATHODE using the ``Baseline" (left) and ``Baseline+$\Delta R$" (right) feature sets. Results are shown without domain adaptation (top) and with domain adaptation using the classification (middle) or regression (bottom) task.}
    \label{fig:cathode spectra}
\end{figure*}

\begin{figure*}[p]
    \centering
    \includegraphics{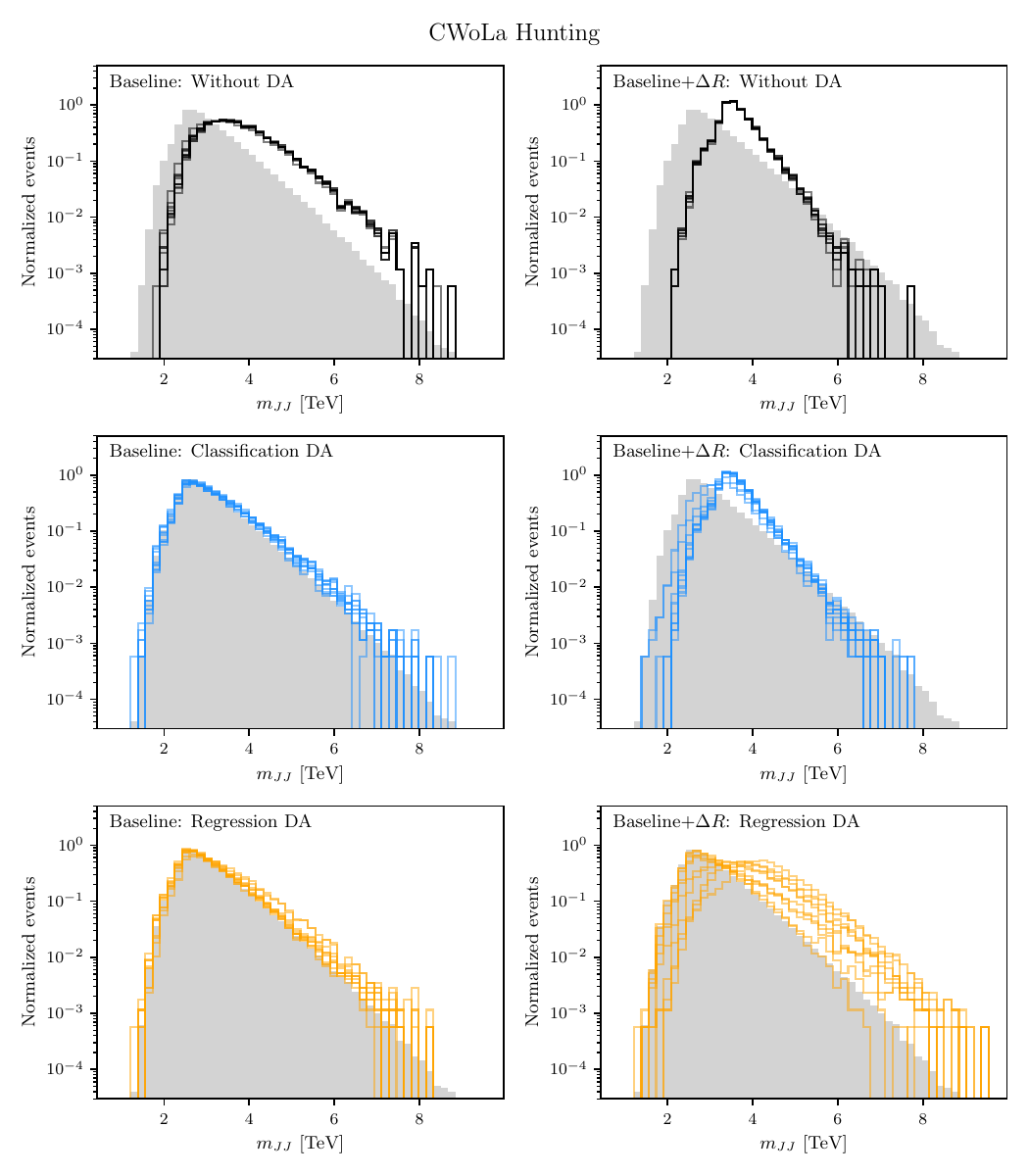}
    \caption{Full set of spectra for ten runs for CWoLa Hunting using the ``Baseline" (left) and ``Baseline+$\Delta R$" (right) feature sets. Results are shown without domain adaptation (top) and with domain adaptation using the classification (middle) or regression (bottom) task.}
    \label{fig:cwola spectra}
\end{figure*}

In Figs.~\ref{fig:IAD results}, \ref{fig:cathode results} and \ref{fig:cwola results}, we visualized the sculpting severity by showing a single example spectrum for each method. In Figs.~\ref{fig:IAD spectra}, \ref{fig:cathode spectra} and \ref{fig:cwola spectra}, we show the full sets of ten spectra to demonstrate the consistency of the results and the variation between runs. In Fig.~\ref{fig:cathode spectra}, we do not show the additional LaCATHODE spectra. These are very stable and can be found in the GitHub repository~\cite{domainadaptationgithub}.

\bibliographystyle{apsrev4-1}
\bibliography{HEPML, other}
\end{document}